\documentclass[aps,prb,showpacs,twocolumn,superscriptaddress]{revtex4-1}

\usepackage{graphicx}
\usepackage{amsmath}
\usepackage{amssymb,gensymb}
\usepackage{wasysym}
\usepackage{textcomp}
\usepackage{color}

\newcommand{\bnen}{\begin{equation}}
\newcommand{\eden}{\end{equation}}
\newcommand{\bean}{\begin{eqnarray}}
\newcommand{\eean}{\end{eqnarray}}

\newcommand{\bna}{\begin{array}}
\newcommand{\eda}{\end{array}}

\begin{document}

\title{Interaction effects in a chaotic graphene quantum billiard}
\author{Imre Hagym\'asi }
\affiliation{Strongly Correlated Systems "Lend\"ulet" Research Group, Institute for Solid State
Physics and Optics, MTA Wigner Research Centre for Physics, Budapest H-1525 P.O. Box 49, Hungary}
\author{P\'eter Vancs\'o}
\affiliation{2D Nanoelectronics "Lend\"ulet" Research Group, Institute  of Technical Physics and 
Materials Science, HAS Centre for Energy Research, Budapest H-1525 P.O. Box 49, Hungary}
\affiliation{Department of Physics, 
University of Namur, 61 rue de Bruxelles, 5000 Namur, Belgium}
\author{Andr\'as P\'alink\'as}
\author{Zolt\'an Osv\'ath}
\affiliation{ Institute of Technical Physics and Materials Science,  HAS Centre for Energy Research, 
Budapest H-1525 P.O. Box 49, Hungary}
\begin{abstract}
We investigate the local electronic structure of a Sinai-like, quadrilateral graphene
quantum billiard with 
zigzag and armchair edges using scanning tunneling microscopy at room temperature. It is revealed 
that besides the $(\sqrt{3}\times\sqrt{3})R30\degree$ superstructure, which is caused by the intervalley scattering, 
its overtones also appear in the STM measurements, which are attributed to the 
Umklapp processes.  We point out that these results can be well understood by taking into account 
the Coulomb interaction in the quantum billiard, accounting for both the measured density of state 
values and the experimentally observed topography patterns. The analysis of the level-spacing 
distribution substantiates the experimental findings as well. We also reveal the magnetic
properties of our system which should be relevant in future graphene based
electronic and spintronic applications.
\end{abstract}
\maketitle
\section{Introduction} 
Quantum dots and quantum billiards have been in the focus of mesoscopic systems for the last three 
decades. The investigation of irregular shaped quantum billiards revealed how the classicaly 
chaotic behavior manifests in their energy spectrums.\cite{rmt1,rmt2,rmt3,rmt4}
Although many properties of quantum dots with two-dimensional electron gas are now well understood,  
the appearance of new two-dimensional materials has renewed the interest in such systems both 
experimentally and theoretically. 
A paradigmatic example is graphene, the two-dimensional honeycomb lattice of carbon atoms,  which 
has been an actively researched area  
since its discovery in 2004.\cite{Novoselov666} Recent developments of nanofabrication and growth 
techniques make it now possible to create nanostructures with well-defined crystallographic 
edges.\cite{Tapaszto:litography} Since they 
serve as building blocks of future nanodevices, it is crucial to understand their properties.
As the Hamiltonian is different and various edge configurations can be present, it is not trivial how 
these effects alter the properties of quantum dots. An illustrative example is a classicaly chaotic 
neutrino billiard, investigated by Berry and Mondragon,\cite{Berry53} where the energy spectrum turned out 
to be governed by the Gaussian Unitary Ensemble (GUE) instead of the Gaussian Orthogonal Ensemble (GOE) as one 
would naively expect, since time-reversal symmetry seems to be present at first glance. The 
Hamiltonian of graphene is similar to that of the neutrino billiards, 
however, the various edge types can lead to further effects.\cite{Jurgen:prl2009,Jurgen:prb2011} 
This is the reason why special 
attention has been paid to graphene based 
billiards.\cite{Jurgen:prl2009,Jurgen:prb2011,huang:chaos,Hamalainen:prl2011,Ponomarenko356,
Rycerz:prb2012,Rycerz:prb2013,Ying:prb2014,Jolie:prb2014,Ramos:prb2014,Ramos:prb2016} It has been 
shown that in the absence of 
intervalley scattering, which case is identical to the neutrino billiard, the level-spacing 
distribution follows GUE statistics.\cite{Jurgen:prl2009,Jurgen:prb2011} However, when the valleys 
are not independent the time-reversal symmetry is restored, the level-spacing distribution obeys GOE statistics. 
\cite{Jurgen:prl2009,Jurgen:prb2011}
\par Although many properties of bulk graphene can be understood within a noninteracting picture, 
interaction effects can become important if one considers finite samples. Even a weak 
electron-electon interaction can lead to drastic effects, for example, in zigzag graphene 
nanoribbons the paramagnetic ground state becomes unstable against magnetic 
ordering.\cite{louie:prl2006,Louie:GW,PhysRevLett.101.096402,Rossier:HF,Yazyev:prl2008,
MacDonald:prb2009,Feldner:prl2011,Schmidt:QMC,Hagymasi:2016} This is due 
to the fact that the zigzag edge states form a flat band at the Fermi energy and the corresponding 
large density of states is not favorable energetically in the presence of the interaction, 
therefore a symmetry-breaking ground state occurs. The same scenario can happen in graphene quantum 
dots possessing zigzag edges.\cite{nanodisks:prb2007,Feldner:prb2010}
\par In this paper we examine a quadrilateral shaped graphene quantum dot, which is a 
truncated triangle having three zigzag and one armchair edges. This peculiar shape
resembles  
the theoretically well-studied Sinai billiard.\cite{stock} In this joint experimental and
theoretical 
study, we use scanning tunneling 
microscopy (STM) and perform theoretical calculations to understand the experimental results. Our 
main finding is that the electron-electron 
interaction must be taken into account to 
reproduce the STM images and local density of states (LDOS) measurements which highlight 
the important role of the interactions in graphene quantum dots even at room temperature. The 
results may have implications in the nanoscale engineering of the electronic and magnetic properties 
of graphene based functional surfaces.
\par The paper is organized as follows. In Sec.~II.~A the investigated quantum dot is introduced together 
with the experimental details, while in Sec.~II.~B the applied theoretical methods are described. 
In Sec.~III.~A we examine the properties of the noninteracting 
quantum billiard using the elements of quantum chaos. In Sec.~III.~B we discuss the experimental 
results together with our theoretical findings. Sec.~III.~C presents the magnetic properties of the quantum 
dot based on the theoretical results. Finally, in Sec.~IV.~ our conclusions are presented.

\section{Methods}
\subsection{Experimental details}
\begin{figure}[!ht]
\includegraphics[width=\columnwidth]{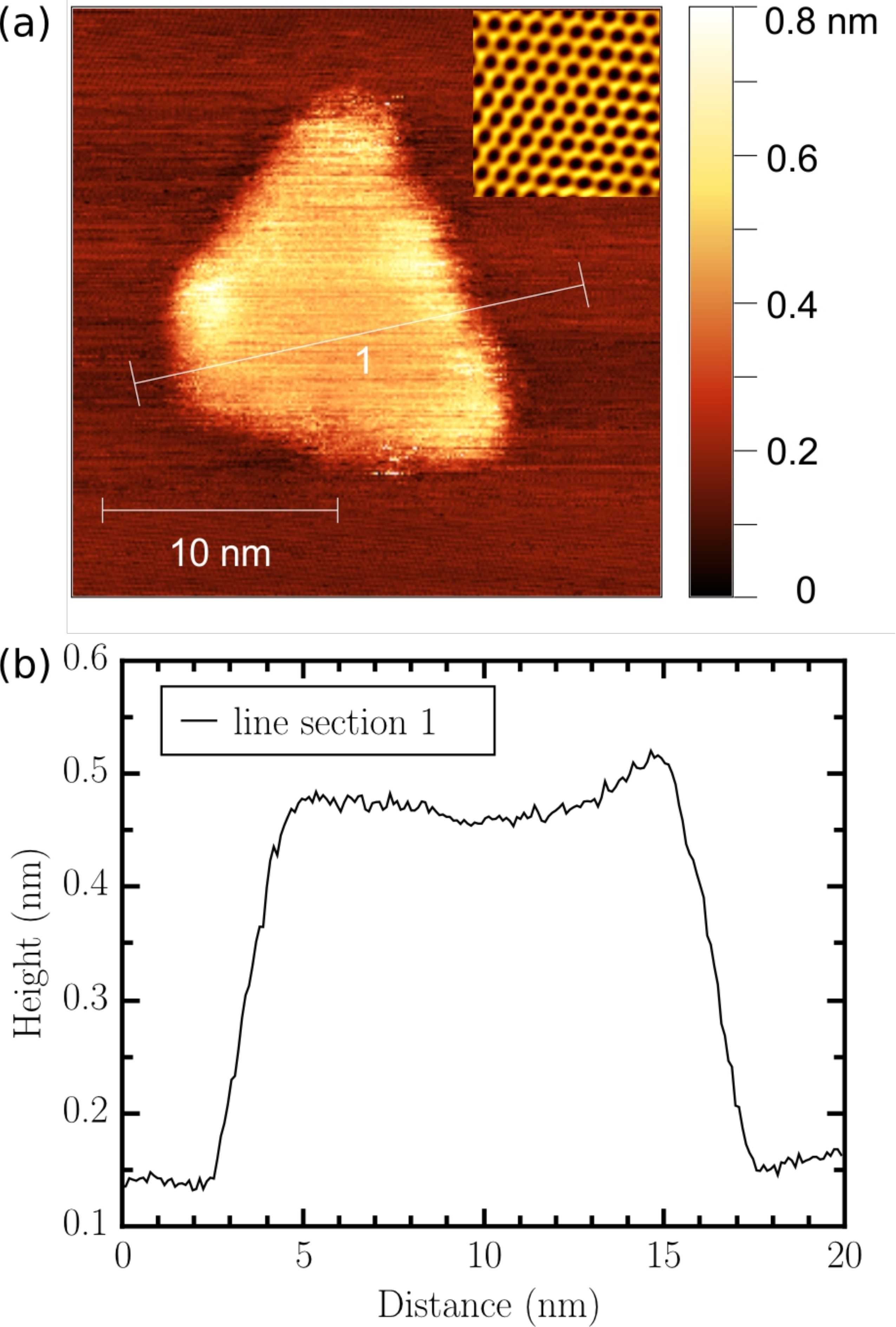}
\caption{(a) STM image of the investigated graphene quantum dot. Tunneling parameters: $U = 200$ 
mV, $I = 1$ nA. The atomic resolution inset image shows the crystallographic orientation in the 
dot. (b) Height profile taken along the line section 1 in (a), showing monolayer thickness. }
\label{fig:dot_geometry}
\end{figure}

Graphene grown by chemical vapour deposition onto electro-polished copper 
foil\cite{Palinkas2016792} was transferred onto highly oriented pyrolytic graphite (HOPG) substrate 
using thermal release tape. An etchant mixture consisting of CuCl2 aqueous solution (20\%) and 
hydrochloric acid (37\%) in 4:1 volume ratio was used. After etching the copper foil, the tape 
holding the graphene was rinsed in distilled water, then dried and pressed onto the HOPG surface. 
The tape/graphene/HOPG sample stack was placed onto a hot plate at 95 $\degree$C. At this 
temperature the tape released the graphene and it could be removed easily. Graphene nanostructures 
were obtained by annealing the graphene/HOPG sample at 650 $\degree$C for two hours in argon 
atmosphere. STM and tunneling spectroscopy (STS) measurements were performed using a DI Nanoscope E 
operating under ambient conditions. The investigated graphene quantum dot is a truncated triangle, 
as shown in Fig.~\ref{fig:dot_geometry} (a). Atomic resolution images obtained on the dot (inset of 
Fig.~\ref{fig:dot_geometry} (a)) reveal that the nanostructure has three zigzag edges and one 
armchair edge. The formation of predominantly zigzag edges is due to the fact that these edges are 
more resistive to thermal oxidation.\cite{Nemes-Incze2010}
\subsection{Theory}
To describe the graphene quantum dot, we use the $\pi$-band model of graphene, where a
lattice site can host two electrons at most with opposite spins. We consider
nearest-neighbor hopping terms only, and model the Coulomb repulsion by using a local
Hubbard interaction term. Thus, we arrive at the Hamiltonian:
\begin{equation}
\label{eq:Hubbard}
 \mathcal{H}=-t\sum_{\langle ij
\rangle\sigma}\hat{c}^{\dagger}_{i\sigma}\hat{c}^{\phantom\dagger}_{j\sigma}+U\sum_i\hat {
n } _ {i\uparrow}
 \hat{n}_{i\downarrow}-\mu\sum_{i\sigma}\hat {n } _ {i\sigma} ,
\end{equation}
where  the summation $\langle ij \rangle$ extends over all nearest-neighbor pairs. The
operator $\hat{c}^{\dagger}_{i\sigma}$ ($\hat{c}^{\phantom\dagger}_{i\sigma}$) creates
(annihilates) an electron at site $i$ with spin $\sigma$ and $\hat{n}_{i\sigma}$ is the
corresponding particle number operator. The nearest-neighbor hopping amplitude is given by
$t$, $U$ is the onsite Hubbard interaction strength and $\mu$ is the chemical potential. Throughout 
the paper we set $t=3
\ {\rm eV}$, which defines the energy scale of the system and tune the chemical potential close to 
the half-filled case to account for the small $p$-doping in the experiment. We 
assume a weak Coulomb
repulsion $U/t=1$, in agreement with previous experiments.\cite{Magda2014} Since the 
Hamiltonian
(\ref{eq:Hubbard}) can be solved exactly for very small systems, we must use
approximations to obtain results for realistic systems. The most commonly used one is the
mean-field approach, whose application is justified for moderate values of
electron-electron interaction in case of graphene nanodisks.\cite{Feldner:prb2010} By neglecting 
the fluctuation terms in the Hamiltonian
(\ref{eq:Hubbard}), we obtain an effective single-particle Hamiltonian
\begin{equation}
 \label{eq:Hamiltonian_mf}
 \mathcal{H}_{\rm MF}=-t\sum_{\langle ij
\rangle\sigma}\hat{c}^{\dagger}_{i\sigma}\hat{c}^{\phantom\dagger}_{j\sigma}+U\sum_{
i\sigma}\langle \hat{n}_{i\bar{\sigma}} \rangle \hat{n}_{i\sigma}-\mu\sum_{i\sigma}\hat {n } _ 
{i\sigma},
\end{equation}
where the unknown electron densities, $\langle \hat{n}_{i\bar{\sigma}}\rangle$ are determined
by using the standard self-consistent procedure and $\mu$ is determined by the conservation of the 
electron number. Having obtained the mean-field solution
one can calculate then the LDOS in the quantum dot as a function of energy  from the Green function:
\begin{equation}
 \rho_i(E)=-\frac{1}{\pi}{\rm Im} \ G_{ii}(E)=-\frac{1}{\pi}{\rm Im}
((E+i0^{+})I-\mathcal{H}_{\rm MF})^{-1}_{ii},
\end{equation}
where $I$ is the identity operator.

\section{Results}

\subsection{Noninteracting case}
\begin{figure}[ht]
\includegraphics[width=\columnwidth]{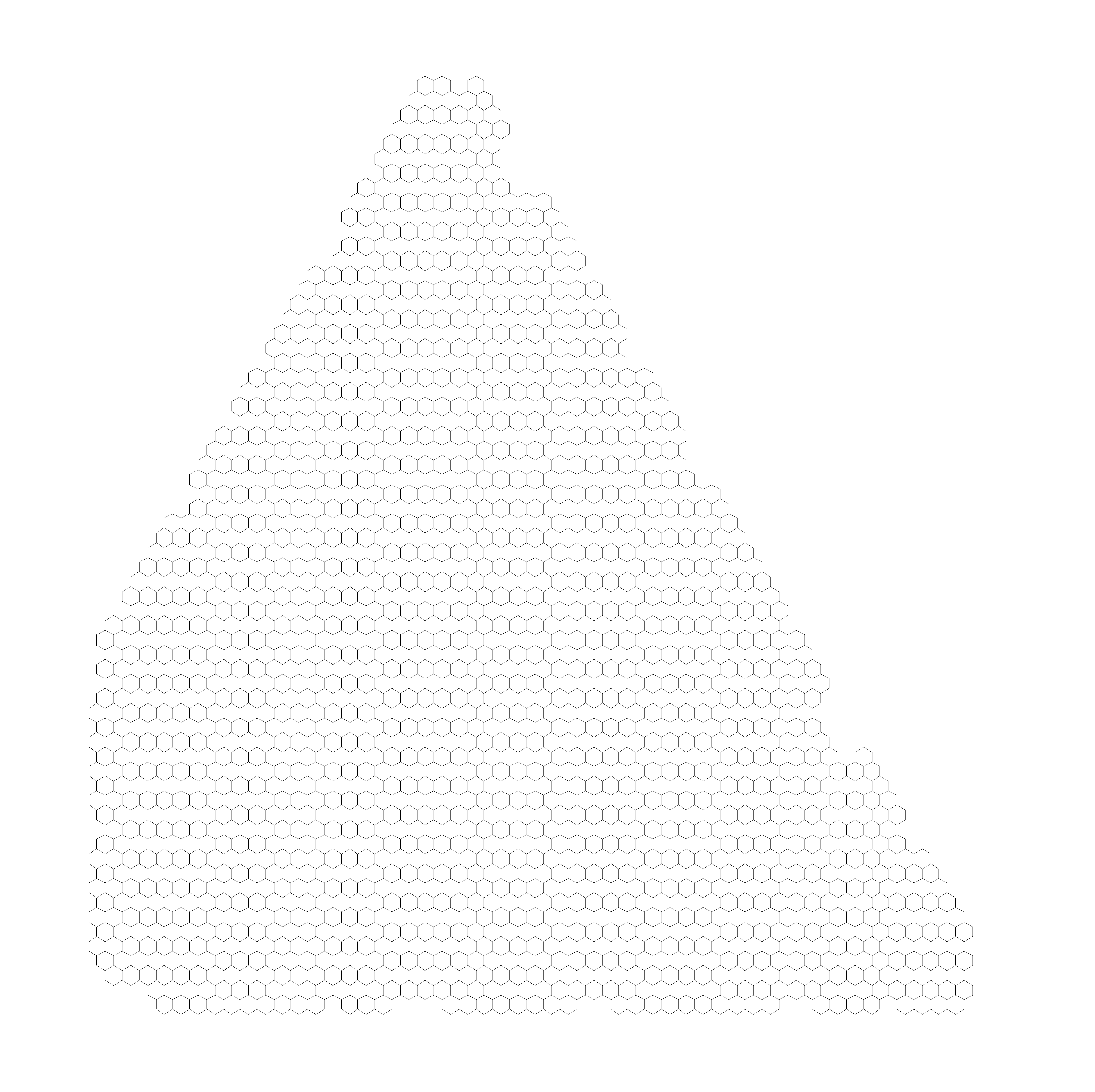}
\caption{The geometry used in the model calculations. The dot consists of 4310 atoms.}
\label{fig:dot_geometry_model}
\end{figure}

\begin{figure}[ht]
\includegraphics[width=\columnwidth]{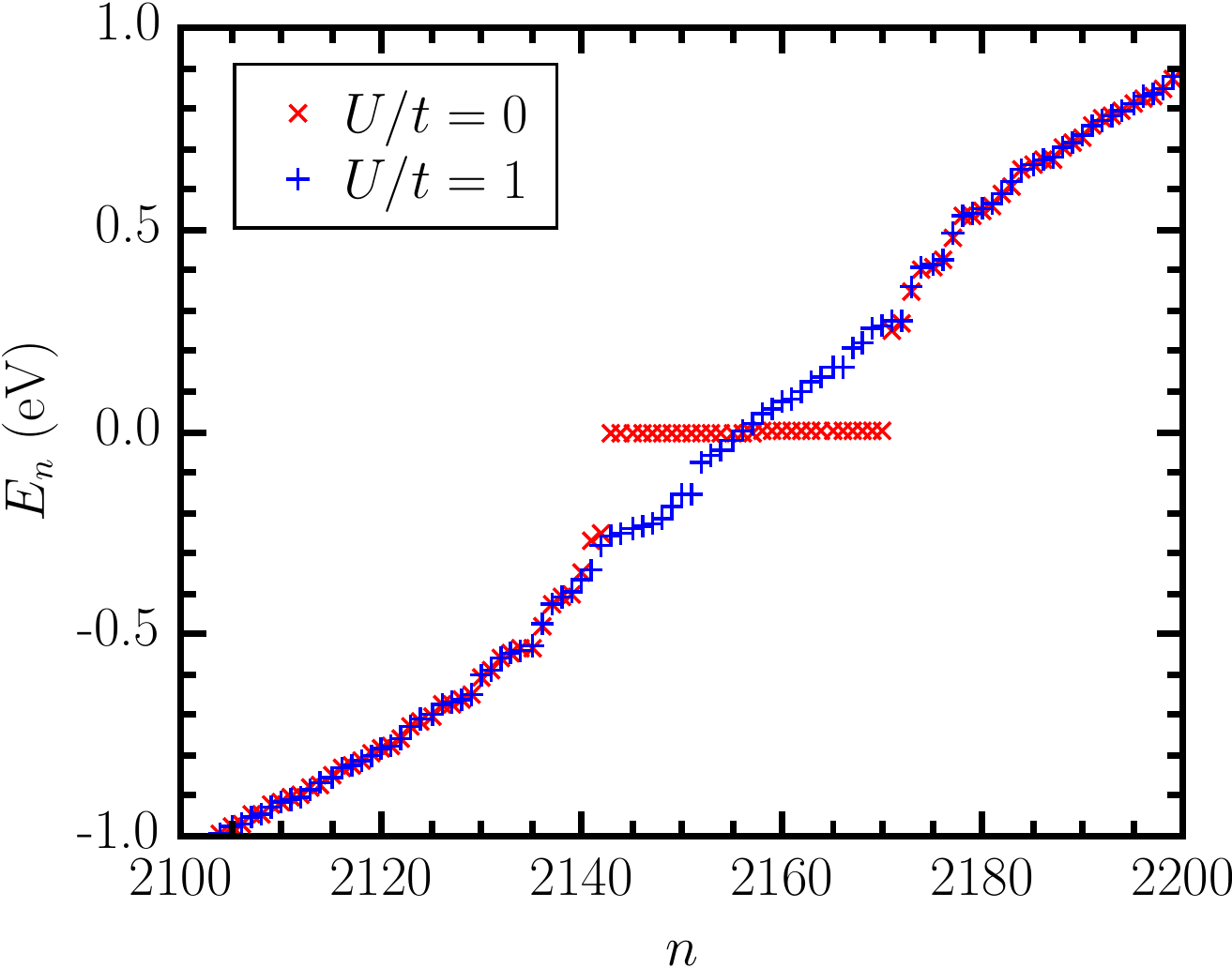}
\caption{The energy spectrum of the quantum dot near the Fermi-energy with and without the Hubbard 
term. The role of the interaction is discussed in Sec.~II.~B. }
\label{fig:energy_spectrum}
\end{figure}

Before diving into the details of the experimental and theoretical results, it is worth 
investigating what we can learn from the noninteracting tight-binding picture.
Therefore, we perform tight-binding 
calculations for this peculiar geometry, a truncated triangle with zigzag and armchair edges, shown 
in Fig.~\ref{fig:dot_geometry} (a). To account for the edge roughness 
and to obtain a more realistic geometry, we removed randomly certain edge atoms as shown in Fig.~\ref{fig:dot_geometry_model}. 
Thus our model contains 4310 atoms. We calculate the energy spectrum by diagonalizing the tight-binding 
Hamiltonian, which is shown in Fig.~\ref{fig:energy_spectrum} ($U/t=0$). 
It is observed that the zero-energy 
states, which correspond to the Fermi level, are multiply degenerate. This may not surprise us if 
we recall that zigzag-shaped nanodisks also have degenerate zero-energy states, which are 
attributed to the presence of edge-localized states.\cite{nanodisks:prb2007} It is worth mentioning 
that this degeneracy is 
still present but partially resolved in our case due to the armchair edge and edge defects. 
\par
\begin{figure}[!ht]
\includegraphics[width=\columnwidth]{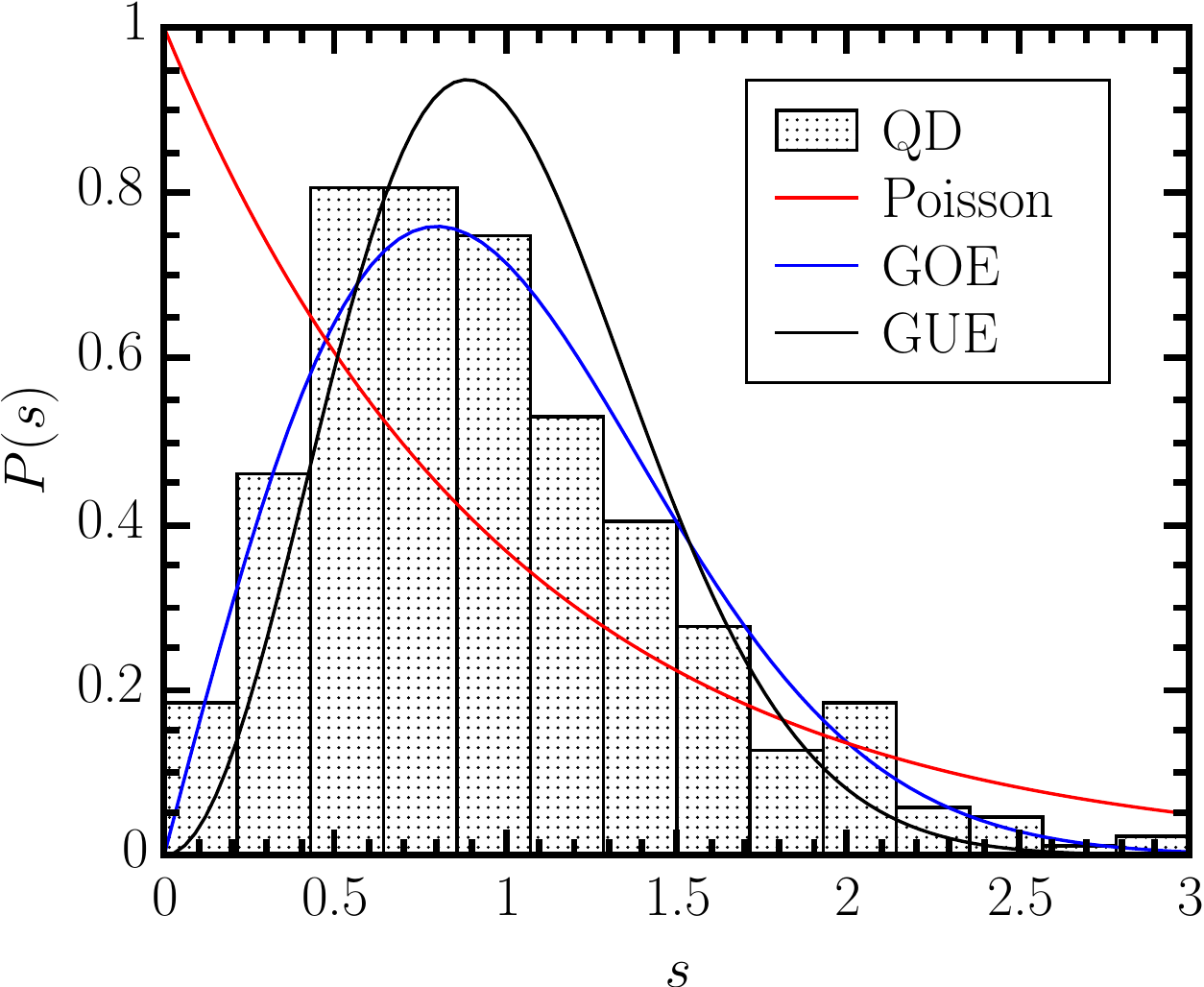}
\caption{The level-spacing distribution obtained from the unfolded spectrum. The solid red, blue 
and black lines correspond to the Poisson, GOE and GUE distributions.}
\label{fig:level_dist}
\end{figure}

 To reveal the properties of integrability, we examine the level-spacing distribution, $P(S)$. 
This quantity reflects the nature of the corresponding classical dynamics of the system and follows 
certain university classes depending on the Hamiltonian.\cite{rmt3,rmt4} By definition, $P(S)dS$ 
gives the 
probability that the distance between the neighboring values of the unfolded spectrum, 
$\langle\rho(E)\rangle(E_{i+1}-E_i)$, lies in the interval $(S,S+dS)$, where $E_i$ are the 
eigenvalues of the tight-binding Hamiltonian and $\langle\rho(E)\rangle$ is the average density of 
states in the interval $(E,E+dE)$. In graphene quantum billiards, we can approximate this quantity by its 
bulk value:
\begin{equation}
 \langle\rho(E)\rangle\approx\rho_{\rm bulk}(E)=\frac{1}{\pi}\frac{\cal A}{(\hbar v_F)^2}|E|,
\end{equation}
where $\mathcal{A}$ is the area of the quantum billiard, $v_F$ is the Fermi velocity. According to our 
previous definition in Sec.~II.~B, $t=\frac{2}{3}\sqrt{3}\hbar v_F/a\approx3 \ {\rm eV}$ and $a$ is 
the lattice spacing 
of graphene. 
In completely integrable systems, the energy levels are uncorrelated, and their level-spacing 
statistics follows Poisson distribution:
\begin{equation}
\label{eq:poisson}
 P(S)=\exp(-S).
\end{equation}
In chaotic systems with (or without) time-reversal symmetry the level-spacing 
obeys the GOE (or GUE) of random matrices:\cite{rmt3,rmt4}
\begin{align}
\label{eq:goe}
 P_{\rm GOE}(S)&=\frac{\pi}{2}S\exp\left(-\frac{\pi S^2}{4} \right),\\
 P_{\rm GUE}(S)&=\frac{32}{\pi^2}S^2\exp\left(-\frac{4 S^2}{\pi} \right).
 \label{eq:gue}
\end{align}
It is worth noting that in integrable systems energy levels lie close to each other, since the 
probability distribution is maximal at $S=0$. In contrast, chaotic systems exhibit level repulsion 
since the corresponding probability densities vanish as $S\to0$, which is often referred as 
avoided level crossings.  
The zero-energy states, which are essentially degenerate,
are due to the zigzag edges and localized on segments of the zigzag boundaries. Since they
would cause an artificial contribution in the level distribution, we consider the 
interval $0.1\lesssim E/t\lesssim1$, which contains approximately 400 eigenvalues.
The level spacing distribution for our noninteracting quadrilateral graphene billiard is shown in Fig.~\ref{fig:level_dist} 
together with the distribution functions Eqs.~(\ref{eq:poisson})-(\ref{eq:gue}) corresponding to the 
university classes. 
In contrast to the equilateral triangle billiard which is an integrable 
system\cite{Rycerz:prb2012} our results clearly indicate that our quantum billiard is a classicaly 
chaotic system.
Moreover, the calculated level-spacings are in a very good agreement with the GOE statistics and 
not with the GUE as one might expect based on the connection with the neutrino billiard. This 
confirms what has been mentioned in Sec.~I that the two valleys are coupled to each other due to the 
armchair edge and the defects on the zigzag edges, which restore the time-reversal symmetry of the 
quantum billiard.

\subsection{Experimental results and their theoretical explanations}
In the previous subsection we have demonstrated that our quantum dot is a chaotic billiard based
on its level-spacing distribution and the corresponding university class is GOE, which
suggests strong intervalley scattering. In what follows we show direct experimental
evidence which supports this scenario and also demonstrate how the electron-electron 
interaction modifies the measured properties of the system.
\par
\begin{figure}[!ht]
\includegraphics[width=\columnwidth]{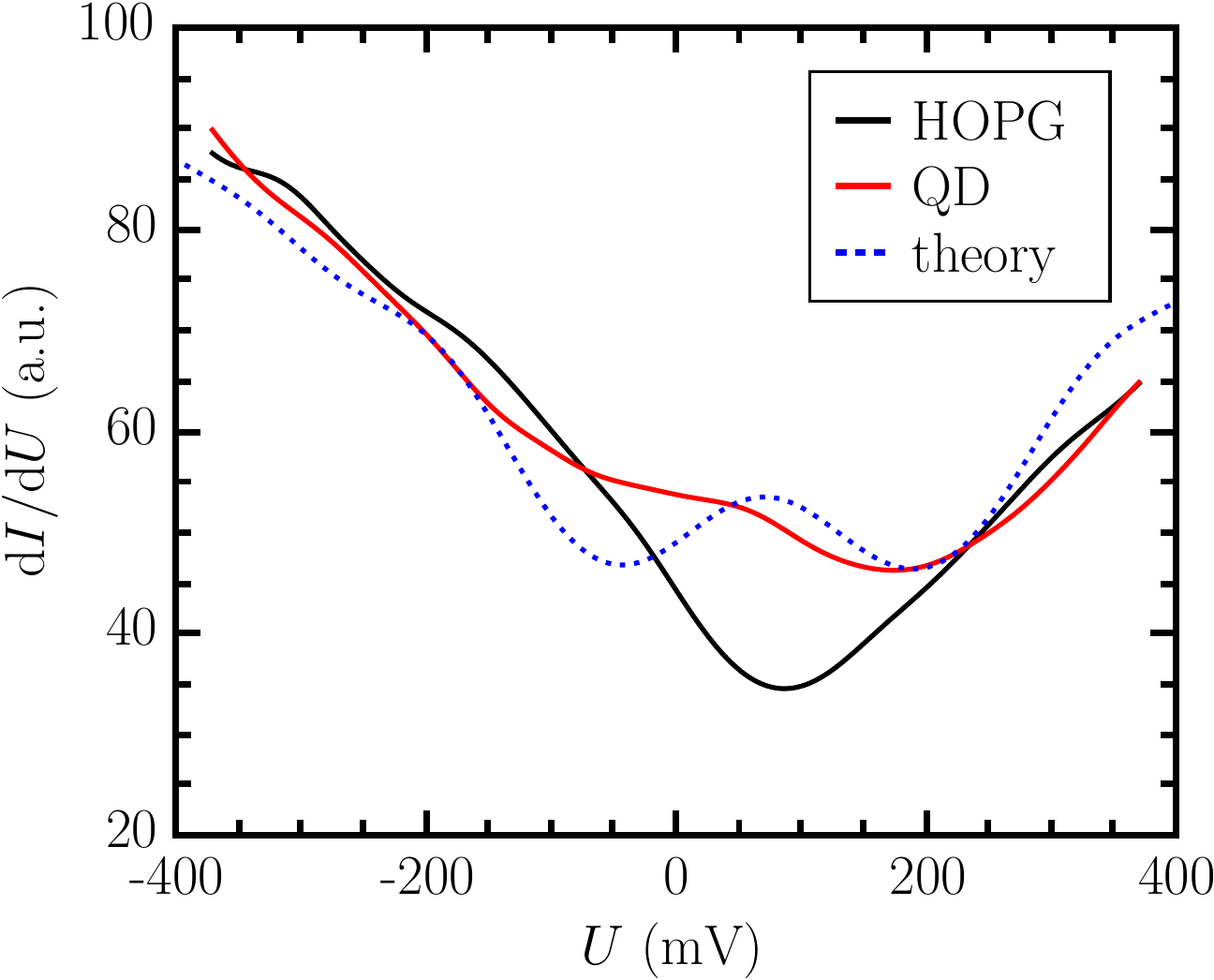}
\caption{The measured LDOS as a function of the voltage, the solid black and red lines correspond 
to the HOPG substrate and the quantum dot, respectively. The dotted blue line represents the 
theoretical result for the LDOS of the quantum dot. In the theoretical calculation a Gaussian broadening $0.1$ eV is 
used to account for the finite temperature. }
\label{fig:dos}
\end{figure}

 First of all, we discuss the measured ${\rm d}I/{\rm d}V$ spectra which correspond to the LDOS 
of the systems. The 
experimental results measured in the middle of the quantum dot and on the HOPG substrate are shown 
in Fig.~\ref{fig:dos}.
The measured HOPG LDOS exhibits the known linear behavior, but it is asymmetric with 
respect to the 
Dirac point and has a finite amount of $p$-doping (80 meV)  due to impurities in the system. 
\begin{figure*}[!htb]
\includegraphics[scale=0.45]{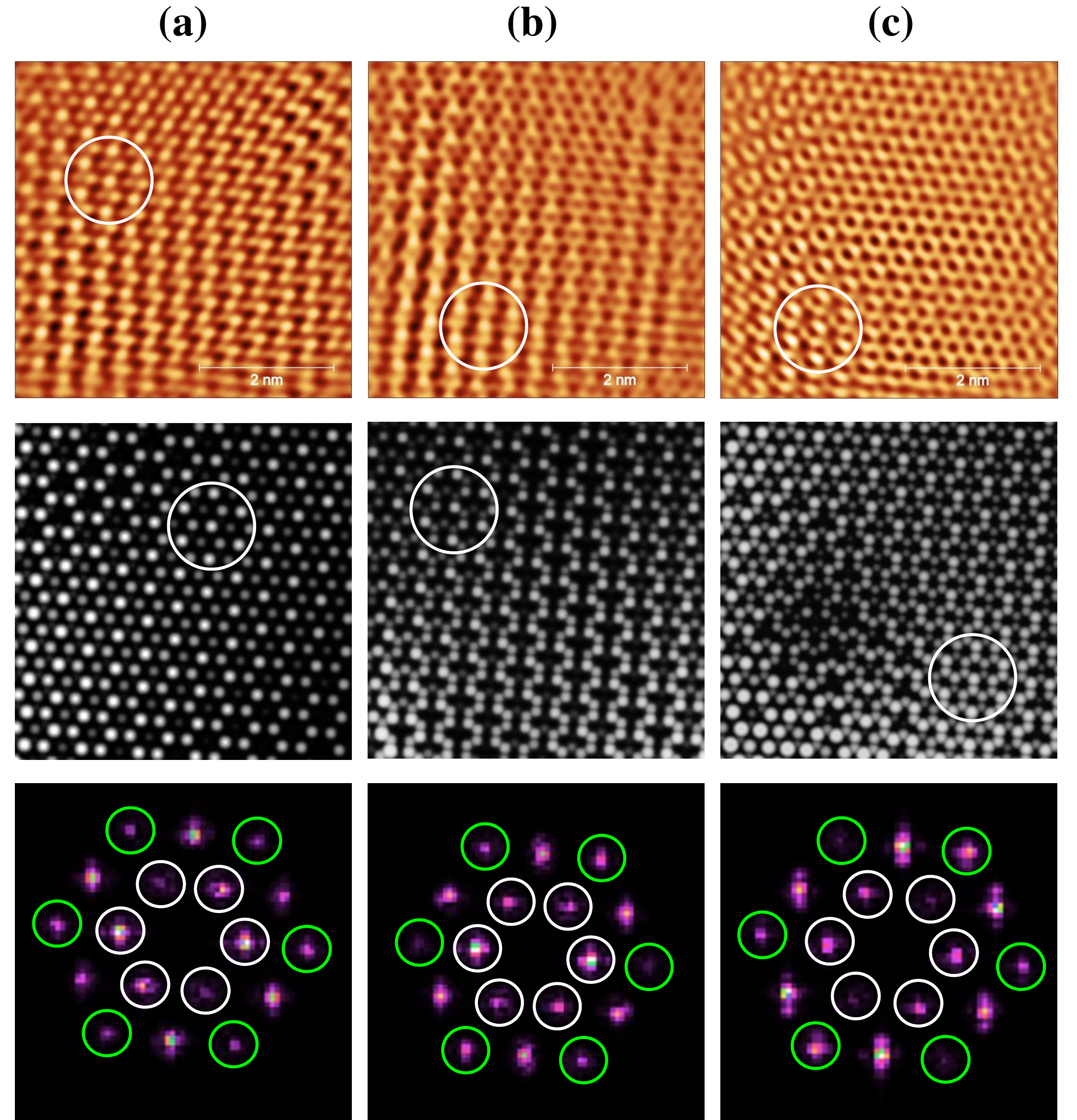}
\caption{The topographic STM images within the quantum dot. Each  column corresponds to different 
voltages where the measurement was done: (a), (b) and (c) belong to $U=100$ mV, 25 mV and -50 mV, 
respectively. The first row shows the measured images, the second one contains the simulated images, 
while  
the third row displays the Fourier spectrum of the measured images. The spots inside the 
white circles  in the Fourier spectrum are the components of the 
$(\sqrt{3}\times\sqrt{3})R30\degree$ superstructure, while 
the green circled spots are their overtones. The non-circled spots correspond to the periodicity of 
the atomic lattice. }
\label{fig:STM_simulation}
\end{figure*}
It is easily seen that the 
quantum dot exhibits similar behavior to the HOPG far from the Fermi energy, however, higher LDOS 
appears near the Dirac point. If we recall the energy spectrum of the noninteracting system in 
Fig.~\ref{fig:energy_spectrum}, we observe that a gap $\Delta\sim0.2$ eV is opened, which is not 
visible in the measured LDOS. 
(We also investigated the role of the next-nearest-neighbor (NNN) hopping term on the 
spectrum of the quantum dot, since it may result in similar effects\cite{PhysRevB.82.045409}. We found that the inclusion of NNN hopping lifts the degeneracy at the Fermi 
level, but the observed band gap quantitatively remains unchanged. See Appendix for details.)
This is the first indication that the noninteracting model does not 
describe the dot accurately.
We know that such highly degenerate levels near 
the 
Fermi-energy are very sensitive to even weak electron-electron interaction, just like in graphene 
nanoribbons with zigzag edges.
Switching on the interaction, we can see from 
Fig.~\ref{fig:energy_spectrum} that the energy levels near the Fermi-energy are no longer 
degenerate even for a small value of the Hubbard-$U$, $U/t=1$, while the high-lying states 
are not modified. (Note that this weak interaction does not alter the energy levels, where the 
level-spacing distribution was taken.) Thus, the obtained spectrum with electron-electron 
interaction is much 
closer to what we expect from the LDOS 
measurements in Fig.~\ref{fig:dos}. 
In order to compare the results we also plot the calculated LDOS inside the quantum dot 
taking into account the finite $p$-doping (Fig.~\ref{fig:dos}).
The measured and the calculated 
values are in good agreement, however, some discrepancy is observed slightly below the Fermi 
energy, where the calculated minimum does not appear.
This can be understood considering that the LDOS of HOPG is asymmetric, therefore its slope below the Fermi energy is larger than 
above it. Since the HOPG also contributes to the tunneling current, their larger LDOS values can 
vanish the calculated LDOS minimum of the dot below the Fermi energy.
\par  In the following we discuss the measured topographic STM images. In 
Fig.~\ref{fig:STM_simulation} we can see superstructure patterns within the quantum dot measured at various voltages.
Regarding the voltages, $U=100$, 25 and -50 mV, where the STM images were  taken, and the 
energy spectrum in Fig.~\ref{fig:energy_spectrum}, then it is evident that the noninteracting tight-binding 
description fails. Since these energy values lie in the gap there is no way to account for the different  STM images. 
In contrast, Fig 6 shows clearly different superstructure patterns. 
In Fig.~\ref{fig:STM_simulation} (a) the well-known 
$(\sqrt{3}\times\sqrt{3})R30\degree$ superstructure (marked by white circle) appears 
on the top of the atomic structure. This is more spectacular from the Fourier spectrum in 
Fig.~\ref{fig:STM_simulation} (a), where the components of the superstructure are clearly seen (marked by white circles). Since this structure originates from the intervalley 
scattering between the $K$ and $K'$ valleys due to the presence of the 
armchair edge,\cite{fukuyama:prb2006,enoki:prb2010} it confirms our previous theoretical findings 
based on the level-spacing distribution (Fig.~\ref{fig:level_dist}) where the intervalley 
scattering was predicted by the GOE distribution.
In Fig.~\ref{fig:STM_simulation} (b)-(c) besides the 
$(\sqrt{3}\times\sqrt{3})R30\degree$ superstructure, stripes parallel to armchair edge and rings 
are observed in the measured STM images. In order to simulate these distinct STM images, we consider a finite 
electron-electron interaction using the procedure described in Sec.~II.~B. 
With the help of the calculated LDOS of the interacting system, we simulate the STM images 
 with the simple Tersoff-Hamann approximation.\cite{tersoff-haman:prb}
Taking into account the electron-electron interaction, we can quantitatively reproduce 
the measured STM images, including the $(\sqrt{3}\times\sqrt{3})R30\degree$ superstructure in 
Fig.~\ref{fig:STM_simulation} (a), the stripes in Fig.~\ref{fig:STM_simulation} (b), and the rings 
in Fig.~\ref{fig:STM_simulation} (c) at the different voltages. If we take a closer look to the measured Fourier spectra, 
it can be easily seen that the Fourier components of the $(\sqrt{3}\times\sqrt{3})R30\degree$ 
superstructure are always present (inner hexagon, components marked with white circles). Moreover, 
their overtones (marked with green circles) appear with various amplitudes which 
results in the formation of different, more complex patterns in the topography images. 
The emergence of the overtones measured by STM can be attributed to the
Umklapp scattering processes, where an electron is scattered outside
the first Brillouin Zone (BZ). Thus, in our quantum dot besides the usual scattering between $K$ 
and $K'$ valleys in first BZ ($\boldsymbol{K-K'=q}$), 
electrons also scatter from the first BZ $K$ valley to the second BZ $K'$ valley ($\boldsymbol{K-K'=q+G}$). 
This latter is equal to the shift of the $K'$ point back to the first BZ by a reciprocal
lattice 
vector $\boldsymbol{G}$. 
Beacuse the Umklapp processes are absent in the
noninteracting case and they appear if electron-electron
interaction is present in the half-filled Hubbard model,\cite{legeza:kdmrg} it serves as
a further evidence for the presence of electron-electron interaction in our system.
\subsection{Magnetic properties}
\begin{figure}[!ht]
\includegraphics[scale=0.12]{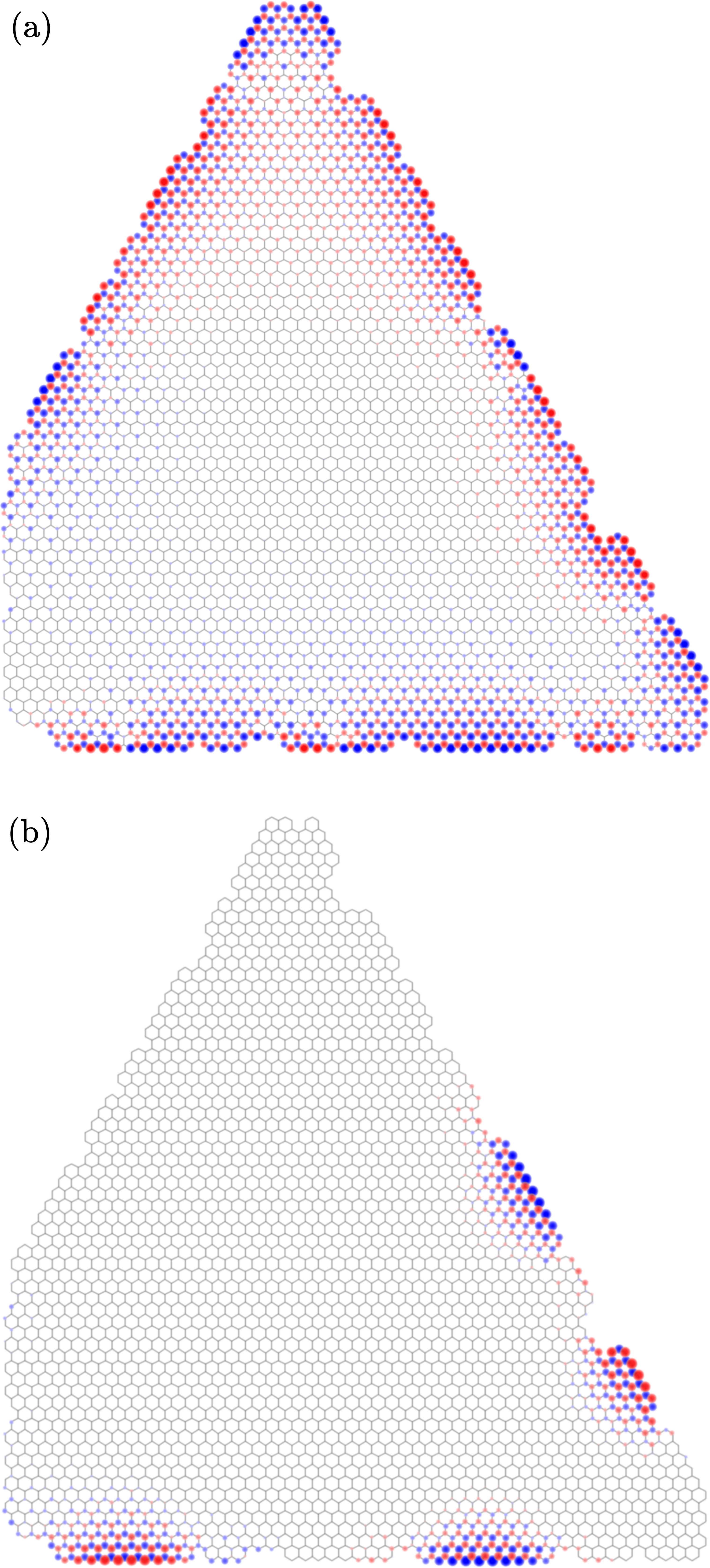}
\caption{Magnetic moments in the quantum dot for $U/t=1$. Panels (a) and (b) correspond to the half-filled case and 80 meV $p$-doping, respectively. The blue and red colors encode the 
up and down spins, respectively. The magnitude of the spins depends on the radii of the circles, $r$, as $\exp(-r)$. The largest magnetic moments in both cases are $S_z\sim0.12$. }
\label{fig:magnesseg}
\end{figure}

Previously, we  demonstrated that the inclusion
of the electron-electron interaction provides a good agreement with the experimental results in terms 
of the 
measured LDOS and topographic STM images. Since our quantum dot is a quite complex system including 
doping and edge defects, 
it is worth revealing the effects of the interaction expicitly by considering the magnetic
moments at the lattice sites. It has been mentioned earlier that the paramagnetic state becomes 
unstable 
against magnetic ordering if zigzag edges are present in
the system,\cite{Feldner:prb2010,Hagymasi:2016,PhysRevB.94.245146} 
however, it also known that the presence of doping can vanish the appeared magnetic 
moments.\cite{PhysRevB.85.075431} 
To clarify the presence of the magnetism in our system we calculated the magnetic moments by taking 
into account the measured $p$-doping (80 meV). We found that magnetic moments are still present in 
the doped system (Fig.~\ref{fig:magnesseg} (b)), although their values are smaller than in the 
half-filled 
case (Fig.~\ref{fig:magnesseg} (a)). 
Quantitatively, the average momentum of  up- and down-spin electrons in the doped quantum dot is 
decreased to one fourth of the undoped value. This reduction can be seen also 
in Fig.~\ref{fig:magnesseg} (b), where the magnetic moments disappeared along the shortest zigzag 
side of the triangle and are localized only at the longer zigzag edge segments. We note that for larger doping values (120 
meV) the magnetic moments totally vanish at the edges, which is consistent with previous 
findings.\cite{PhysRevB.85.075431}
Our calculations indicate that only the zigzag edges are magnetic and no magnetization occurs 
along the armchair edge and around the defects in the zigzag edges. It
is in agreement with the \emph{a priori} expectations based on nanoribbons, since armchair nanoribbons are 
not magnetic, and edge defects in zigzag ribbons suppress the 
magnetism.\cite{Magda2014,PhysRevB.82.085425} This follows from the fact that here
the sublattice symmetry is restored, while zigzag edges
contain atoms only from one sublattice. We also observe spin-collinear domain 
walls\cite{PhysRevLett.100.047209} along the edges due
to presence of the defects. Overall, we can conclude that the magnetism can be present in the system 
even at the presence of small doping and edge defects based on the mean-field calculations. Our analysis above gives further insight into the role of the electron-electron interaction 
in graphene quantum dots.

\section{Conclusions}
We carried out a joint experimental and theoretical investigation to analyze the properties of a 
quadrilateral shaped graphene quantum billiard. First, we examined the quantum billiard in 
a tight-binding picture to grasp the properties of integrability. It turned out that the billiard is 
chaotic, moreover, its level-spacing distribution agree well with the GOE distribution. This 
indicates strong intervalley scattering, which also manifests in the appearance of the 
$(\sqrt{3}\times\sqrt{3})R30\degree$ superstructure observed in the STM images measured under ambient conditions. 
We pointed out that 
by taking into account the interaction among the electrons, we were able to elucidate the 
spectroscopy (LDOS) measurements and STM images simultaneously. Using the mean-field approximation 
of the Hubbard model we reproduced the $(\sqrt{3}\times\sqrt{3})R30\degree$, stripe and 
ring superstructures as well. It was also revealed that the latter two structures appear due 
to the overtones of the Fourier components of $(\sqrt{3}\times\sqrt{3})R30\degree$ superstructure, 
which also confirms the importance of the interaction even at room-temperature.
Furthermore, we showed that the weak electron-electron interaction does not alter the higher energy levels ($>0.2$ eV), only the ones near the Dirac point (lifted degeneracy). Thus the chaotic nature and the time-reversal symmetry of the graphene billiard is preserved.
 We also 
demonstrated that edge-magnetism appears along the zigzag edges, and the magnetic moments disappear 
inside the quantum dot. The ability of tailoring the electronic and magnetic properties of 
graphene quantum dots can open new avenues for information coding at nanoscale. 

\acknowledgements{I.~H.~ and P.~V.~ contributed equally to this work. The research leading to these 
results has received funding from the People Programme (Marie Curie Actions) of the European Union’s 
Seventh Framework Programme under REA grant agreement No.~334377 and “Lend\"ulet” programme of the Hungarian Academy of Sciences grant LP2014-14. Financial support from the 
National Research, Development and Innovation Office (NKFIH) through the OTKA Grant Nos.~K119532 and 
K120569 are acknowledged.}

\appendix*
\section{The role of next-nearest-neighbor hopping}
\begin{figure}[!ht]
\includegraphics[width=\columnwidth]{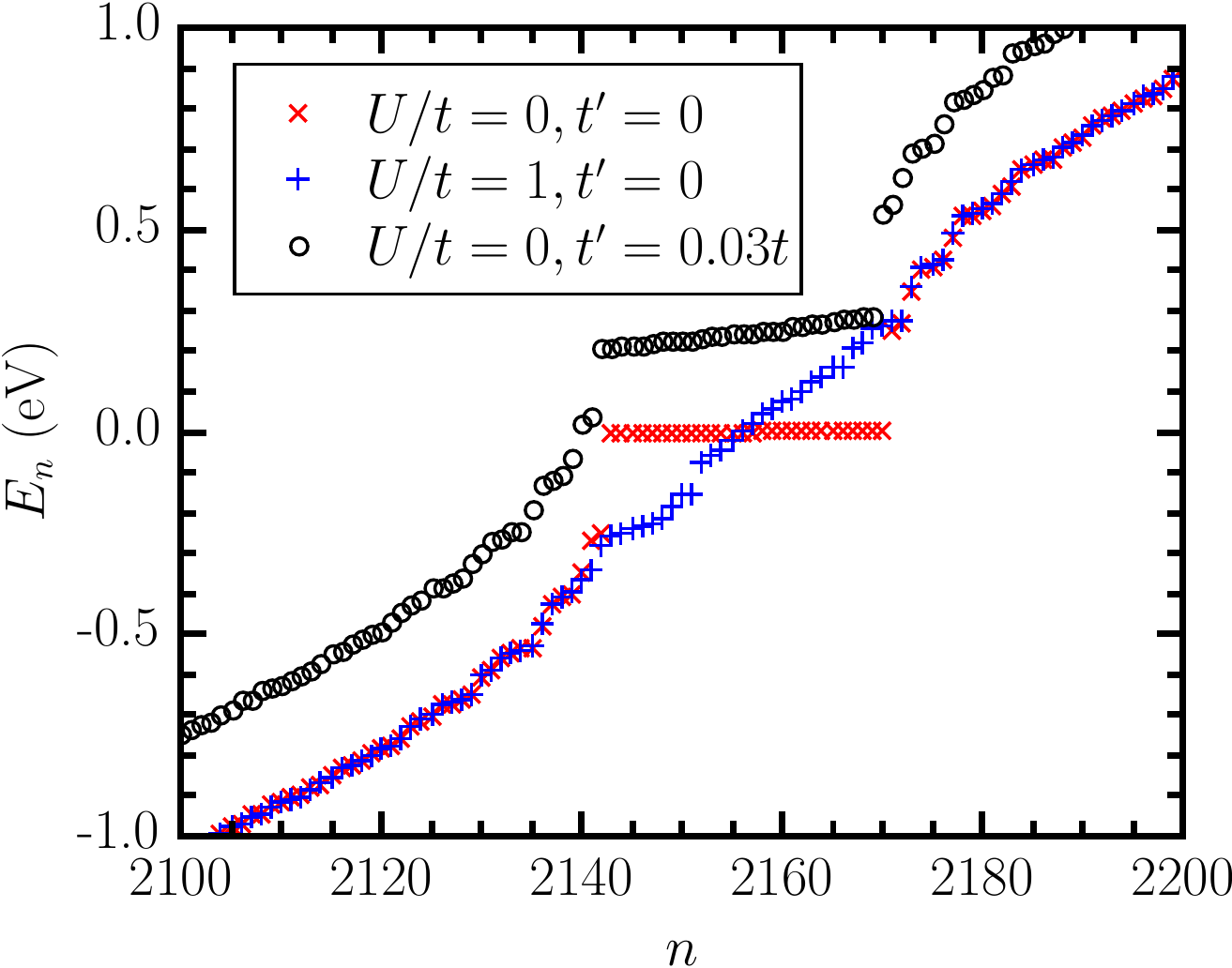}
\caption{ The energy spectrum of the quantum dot near the Fermi-energy with and without NNN 
hopping terms. For better comparison, the result for the interacting case is shown too. }
\label{fig:energy_spectrum_app}
\end{figure}

In order to understand the lack of energy gap in the ${\rm d}I/{\rm d}V$ spectra and the role of 
the interaction it is important to consider other effects that may alter the energy spectrum. The 
most important candidate is the next-nearest-neighbor (NNN) hopping, since it is known to break 
the electron-hole symmetry and shifts the dispersionless edge states away from the zero 
energy.\cite{PhysRevB.82.045409} Therefore, we study the following Hamiltonian:
\begin{equation}
\label{eq:Hubbard_NNN}
 \mathcal{H}=-t\sum_{\langle ij
\rangle\sigma}\hat{c}^{\dagger}_{i\sigma}\hat{c}^{\phantom\dagger}_{j\sigma}-t'\sum_{\langle\langle 
ij
\rangle\rangle\sigma}\hat{c}^{\dagger}_{i\sigma}\hat{c}^{\phantom\dagger}_{j\sigma} +U\sum_i\hat {
n } _ {i\uparrow}
 \hat{n}_{i\downarrow},
\end{equation}
where $\langle\langle ij\rangle\rangle$ denotes the summation over all NNN pairs. For the value of 
$t'$ we use $t'=0.1$ eV from the tight-binding fit to experimental 
data.\cite{PhysRevB.76.081406,RevModPhys.81.109} To reveal the effect of NNN hopping, we calculated 
the energy spectrum of the Hamiltonian (\ref{eq:Hubbard_NNN}) in the noninteracting case, which is 
shown in Fig.~\ref{fig:energy_spectrum_app}.
It is obviously seen that the NNN hopping lifts the degeneracy of the edge states, and they are 
slightly shifted towards higher energies. However, they still lie very close to each other, and the 
gap $\Delta\sim0.2$ eV in the spectrum hardly decreases. In contrast, the interaction completely 
removes the degeneracy of the zero-energy states and the gap disappears in the spectrum. 
\bibliography{paper_graphene} 

\end{document}